\documentclass[runningheads]{llncs}

\usepackage[T1]{fontenc}

\usepackage{bbding}
\usepackage{graphicx}
\usepackage{cite} 
\usepackage{csquotes} 

\usepackage[hyphens]{url} 
\usepackage[hidelinks]{hyperref}
\usepackage{orcidlink}
\usepackage{color}

\begin{document}
\title{IMoG - a methodology for modeling future microelectronic innovations}
\author{Oliver Klemp\orcidlink{0000-0001-7891-4572} \and
	Bernd Westphal\orcidlink{0000-0002-6824-0567} \and
	Stefan Puch
	}
\institute{German Aerospace Center (DLR)\\ Institute of Systems Engineering for Future Mobility \\
	Escherweg 2, 26129 Oldenburg, Germany\\
	\email{\{oliver.klemp, bernd.westphal, stefan.puch\}@dlr.de}\\
	\url{http://www.dlr.de}}

\maketitle

\begin{abstract}
	\textbf{[Context and motivation]} The automotive industry is currently undergoing a fundamental transformation
  towards software defined vehicles.
  The automotive market of the future demands a higher level of automation,
  electrification of the power train,
  and individually configurable comfort functions.
  \textbf{[Question/problem]} These demands pose a challenge to the automotive development cycle, because they introduce complexity by larger and not yet well explored design spaces that are difficult to manage.
  \textbf{[Principal ideas/results]} To cope with these challenges, the main players along the value chain have an increased interest in collaborating and aligning their development efforts along joint roadmaps.
  Roadmap development can be viewed as a field of requirements engineering
  with the goal to capture product aspects on an appropriate level of
  abstraction
  to speed up investment decisions, reduce
  communication overhead and parallelize development activities,
  while complying with competition laws.
  \textbf{[Contribution]} In this paper, we present a refinement of the
  \enquote{Innovation Modeling Grid} (IMoG), which encompasses a methodology, a
  process and a proposed notation to support joint analysis of development
  roadmaps.
  IMoG is focused on the automotive domain, yet there are clear
  potentials for other applications.

\end{abstract}

\section{Introduction}
\label{sec:Introduction}
The automotive industry faces the following challenges, which requires the automotive value chain to adapt to remain competitive \cite{pwc2018}:
The automotive industry has recognized for a long time the demand and potential of highly automated driving and fully autonomous driving.
This autonomous functionality rely, for example, on perceiving the environment, locating the car's position, computing trajectories, planning and navigating routes to controlling the car.
The implied high software complexity poses a major challenge to the automotive industry \cite{softwareAsInnovationDriverCariad, softwareAsInnovationDriverBosch, softwareAsInnovationDriverInfineon}. 
The demand of individualization, e.g., the support for plugging the newest smartphone generations into the car or the individualizing of the air condition systems, is also software driven, underlining the need to focus on software.

The climate change and the departure from combustion engines is a another challenge.
The expected technology shift goes towards the electrification of cars, which represents an easier technology and opens the markets for start-ups.
Considering the easier technology as well as the limited resources (e.g., lithium), the electrification increases the pressure on the market for all members of the automotive value chain.
Finally, the general trend of car manufacturers to move towards new business models, that focus on mobility as a service, poses a major challenge to the whole value chain.

The long established automotive value chain with their long development times of up to 10 years needs to find a way to reduce the development times to remain competitive \cite{pwc2018}.
One way (and our investigated way) to accelerate the development of the automotive value chain is by trying to boost innovation with a public roadmapping approach.
It is assumed, that such a developed public roadmap is created and maintained by an automotive value chain committee.
This committee may include several car manufacturers (Original Equipment Manufacturer (OEM)), several software and hardware component suppliers (also referred to as Tier 1) as well as several semiconductor suppliers (Tier 2).
The committee is public and open for new members to comply with the compliance laws.
This roadmapping approach helps to better understand and communicate future innovations, the required future technologies and the decisions of other partners about their future direction along the value chain.
Based on the communicated directions, the value chain partners can reduce their risks in investment decisions into new technologies and speed up their decision process.
This roadmapping context represents a communication and understanding problem, which is in essence a requirements engineering problem.
However, this context does not represent the typical requirements engineering field:
The public roadmapping does not resemble a product as in the classical case between a corporation and a customer.
It focuses on the pre-development specification of an innovation managed in a distributed manner.

Given this roadmapping approach, we investigated the research question of what an appropriate methodology, process and tool would be.
In our opinion a dedicated methodology supported by a process and tailored tooling is required to handle this special and abstract context.
The methodology targets to efficiently represent and model early microelectronic innovations to enable a consistent information transfer along the value chain on public roadmaps.
The methodology shall reduce the start-up time for innovation modeling by pre-structuring the innovation in the sense of advising what type of elements exist and how they relate to each other.
The process shall answer who is doing what with which tool to produce which artifacts and when they are finished.
The tooling shall support as good as possible the above mentioned process and methodology.

Focusing on this context, we developed a methodology called \enquote{Innovation Modeling Grid} (IMoG) to accelerate the innovation development process along the automotive value chain \cite{Fakih2021}.
The journal article included a preliminary evaluation regarding IMoG's validity, usefulness and adequacy with the result of IMoG being a promising approach.
A definition of IMoG's model elements and a process to answer how to use IMoG remained open.
Therefore, we have further consolidated IMoG, by defining a suitable process with the required roles and by refining IMoG's perspectives and their corresponding model elements.
Additionally, we have explored IMoG in more sophisticated academic innovations.
The contribution of the consolidation of IMoG, its process and its application to the academic innovations will be presented in this article.
The related work is described in Section \ref{sec:relatedwork}.
Afterwards, the IMoG methodology, its process and its roles are presented in Section \ref{sec:IMoG Methodology}.
A discussion of the current state of IMoG and a conclusion complete the article.
\section{Related Work}
\label{sec:relatedwork}
IMoG is designed specifically for its context: Public collaboration between an open number of corporations along the automotive value chain to derive roadmaps for future microelectronic innovations.
We are not aware of any methodologies, processes or tools tackling an identical issue.
Nonetheless, IMoG is inspired by general principles of the requirements engineering field and by methodologies that target innovations.
These methodologies have their specific context, which make them only partially applicable to the context of IMoG.
However, they provide valuable insights on how to tackle the modeling of innovations.

IMoG's matrix shape is inspired by SPES \cite{pohl2012spes}.
The SPES concept of using different perspectives, which handle different aspects of the examined innovation as well as using different abstraction levels are included in IMoG's design principles.
The perspectives and aspects covered differ between IMoG and SPES, however, similar topics are covered.
This SPES concept of perspectives is also used in other methodologies like EAST-ADL \cite{debruyne2004east-adl} and AUTOSAR \cite{autosar}, but both focus on design phases and their artifacts specifically.

Furthermore, IMoG relies on concepts from requirements engineering.
This includes (1) the partition of the problem and solution space \cite{czarnecki2000generative, olsen2015lean} to reduce cognitive load and improve modeling, (2) the use of User Stories and Use Cases \cite{jacobson1993object} to model the user needs in the problem space, (3) the use of feature models \cite{kang1990feature} and subsidiary variants like \cite{czarnecki2000generative}, (4) the partitioning of system descriptions into context, logical architecture, system components and parts similar to AUTOSAR \cite{autosar} and others.

The innovation modeling methodology of Gleirschner et al. \cite{gleirscher_model-based_2014} closely relates to IMoG. 
It uses feature models, requirements, use cases and component models to describe the pre-development phases of automotive innovations.
The contrast to IMoG lies in its context and modeling focus:
Gleirschner et al. emphasize the features and services including behavioral aspects known from systems level design.
In contrast, IMoG emphasizes on the roadmapping context in the committee, which does not focus on providing too much details.
On the other hand, IMoG includes for example the description of strategies on the Strategy Perspective of the stakeholders in the automotive value chain.

\section{IMoG Methodology}
\label{sec:IMoG Methodology}
We developed a methodology called \enquote{Innovation Modeling Grid} (IMoG) to accelerate the innovation development process along the automotive value chain.
The methodology IMoG provides a structure and defines elements to model the problem and the solution space of innovations.
IMoG defines a structure to reduce the time spend on the \enquote{What and how to model?} question and to help the modeler to focus on their innovation instead.
Furthermore, a process and a dedicated tooling supporting the methodology is required to handle the methodology this public roadmapping context.
A dedicated tooling for IMoG is currently in progress, but out of the scope of this article.

Section \ref{sec:imog:overview} presents the design principles of the methodology IMoG.
Section \ref{sec:imog:methodology} describes IMoG's methodology.
IMoG's process is described in Section \ref{sec:imog:process}.

\subsection{Design Principles of IMoG}
\label{sec:imog:overview}

We developed IMoG under the context that an automotive value chain committee creates and maintains a public microelectronic roadmap.
This context has implications on the methodology and thus shaped IMoG's design principles.
These design principles are outlined here:
\begin{itemize}
	\item \textbf{Abstract Innovations}: The public collaboration of the corporations of the automotive value chain for future innovations requires an abstract representation of the innovation to remain beneficial to all participants.
	IMoG models are expected to include fewer details than typical development and engineering models and therefore, complex modeling concepts are left out.
	This includes, for example, the concept of \enquote{Ports} to model communication interfaces and check their consistency.
	However, this does not mean that any kind of detail is too much.
	IMoG is expected to contain sufficient details of the innovation's crucial parts where the highest uncertainty and risk lie.
	Instead of Ports, one communication channel describing sufficient details is recommended.
	\item \textbf{Problem space vs solution space}: The innovation shall be divided into a problem description and a solutions description \cite{czarnecki2000generative, olsen2015lean}:
	The problem space should mostly contain information about the problem with as little information as necessary about the possible solutions.
	The solution space covers then the possible solutions.
	Furthermore, a map between the problem space elements and the solution space elements is necessary for basic tracing.
	Natural language constraints, quality requirements and general conditions complete this tracing by giving the options to add further information.
	In the context of a roadmapping committee, this problem-solution distinction is suitable, because it eliminates the frequently asked question whether a particular \enquote{Function}, \enquote{Block}, \enquote{Requirement} in the IMoG model describes the target state or the actual state and thus helps reducing the thinking overhead.
	\item \textbf{Support of Decomposition / Refinement / Variability}: These three core concepts shall be distinguished whenever needed to maximize usability.
	In the context of public roadmapping the distinction is less important for describing the problem space.
	However it is invaluable to understand and apply these concepts for the solutions and their variants.
	\item \textbf{Omitting behavioral or structural focus}: The focus in context of a microelectronic roadmap lies on the understanding what the problem is and not on which parts are needed to build the innovation.
	Thus structural details are intentionally left out.
	Similarly, the solution description focuses on how solutions can be implemented, on their properties and on their variants: Behavioral details are kept abstract for the the blocks and variants.
	\item \textbf{Abstraction Levels and Perspectives}: The concept of abstraction levels and perspectives shall help with the separation of concerns as well as with the support of filtering mechanisms to hide temporally unneeded details.
\end{itemize}

\subsection{Innovation Modeling Grid Methodology}
\label{sec:imog:methodology}

\begin{figure}
	\includegraphics[width=\linewidth]{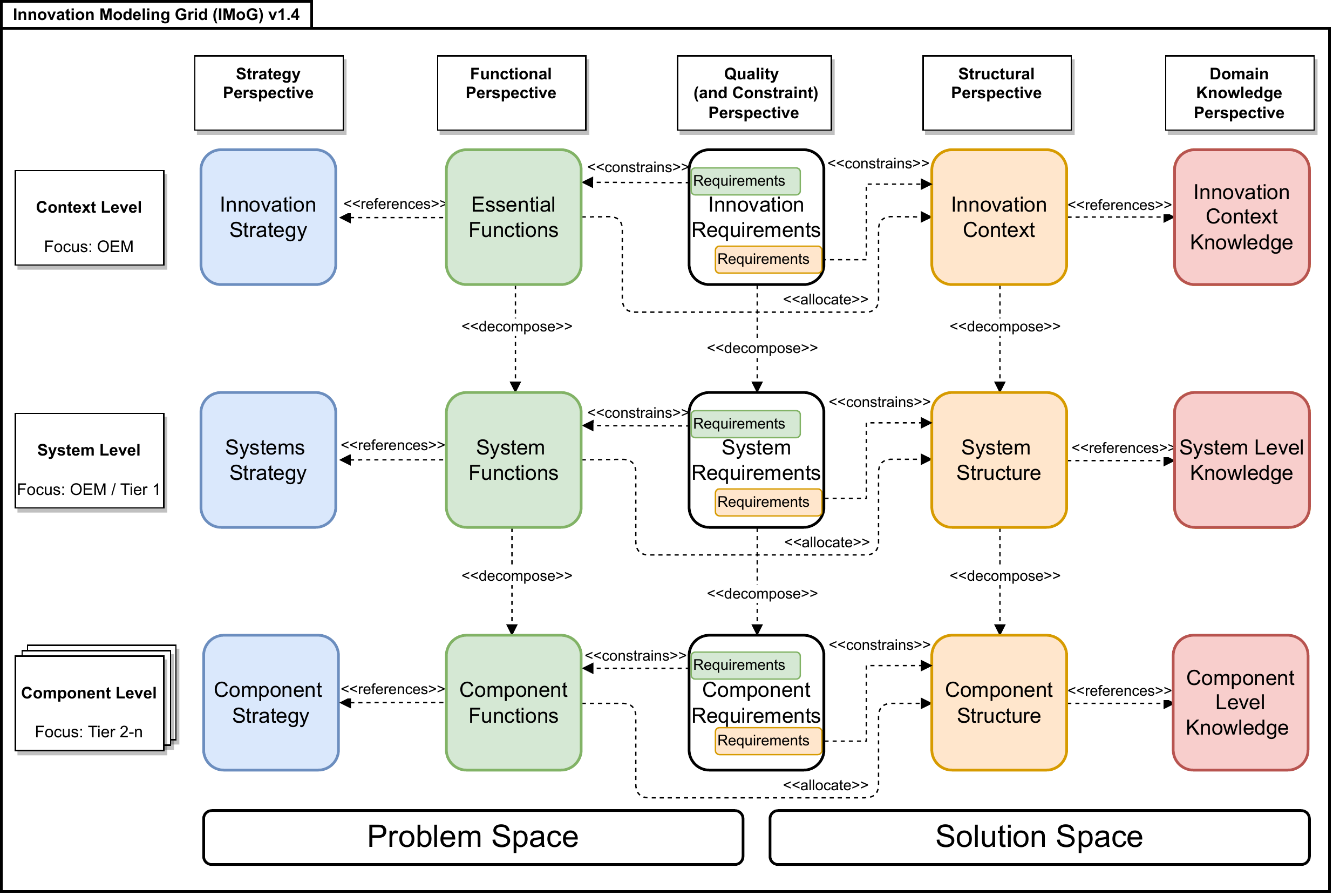}
	\caption{IMoG version 1.4.
	It contains three abstraction levels (rows) and five perspectives (columns).
	Each perspective and abstraction level is interconnected with its neighbor cell.}
	\label{fig:imog1.4}
\end{figure}

The Innovation Modeling Grid (IMoG) is depicted in Figure \ref{fig:imog1.4}.
IMoG forms a matrix with rows and columns.
Each row represents an abstraction level, which can be understood as separating and designing the details of the innovation at different detail levels.
IMoG currently proposes three abstraction levels:
\begin{itemize}
	\item The \textbf{Context Level} is used to describe the innovation as a whole system embedded in its environment.
	\item The \textbf{System Level} describes the system and its parts.
	\item The \textbf{Component Level} describes the components of the system and parts in more detail.
\end{itemize}

Each column represents a perspective that describes a different aspect of the innovation.
IMoG defines two categories of perspectives:
\begin{itemize}
	\item Perspectives of the \textbf{problem space} focus on the description of the problem, ignoring all technical information.
	These include the Strategy Perspective, the Functional Perspective and partially the Quality Perspective.
	\item Perspectives of the \textbf{solution space} focus on how to translate the problem into solution designs. In the context of innovations these solution spaces are kept abstract as the knowledge about the future is only vague.
	The Structural Perspective, the Knowledge Perspective and the other part of the Quality Perspective correspond to the solution space.
\end{itemize}
The IMoG meta model recommends for each perspective a set of model elements.
The corresponding details are out of the scope of this article.
Each perspective is presented in the following.
Afterwards, the interconnection (and thus the arrows in Figure \ref{fig:imog1.4}) between the perspectives are described.

\subsubsection{Strategy Perspective}
\label{sec:imog:methodology:strategy}
An innovation's creation usually starts with many discussions, sketches and creative methods.
These discussions are the starting point of the Strategy Perspective.
The role of the Innovation Leader (see Section \ref{sec:imog:process}) -- possibly the initiator in a committee -- takes the outcome and writes the innovation description.
The description targets the innovation strategies, which may contain a vision, rationales, images, goals and diagrams.
These descriptions can contain identifiable elements to allow referencing and tracing.
Additionally, the description contains abstract company strategies and information like their stake in the innovation.
The description and identifiable elements encompass together enough information to start the modeling activities from the other Perspectives.

\begin{figure}
	\centering
	\includegraphics[width=0.85\linewidth]{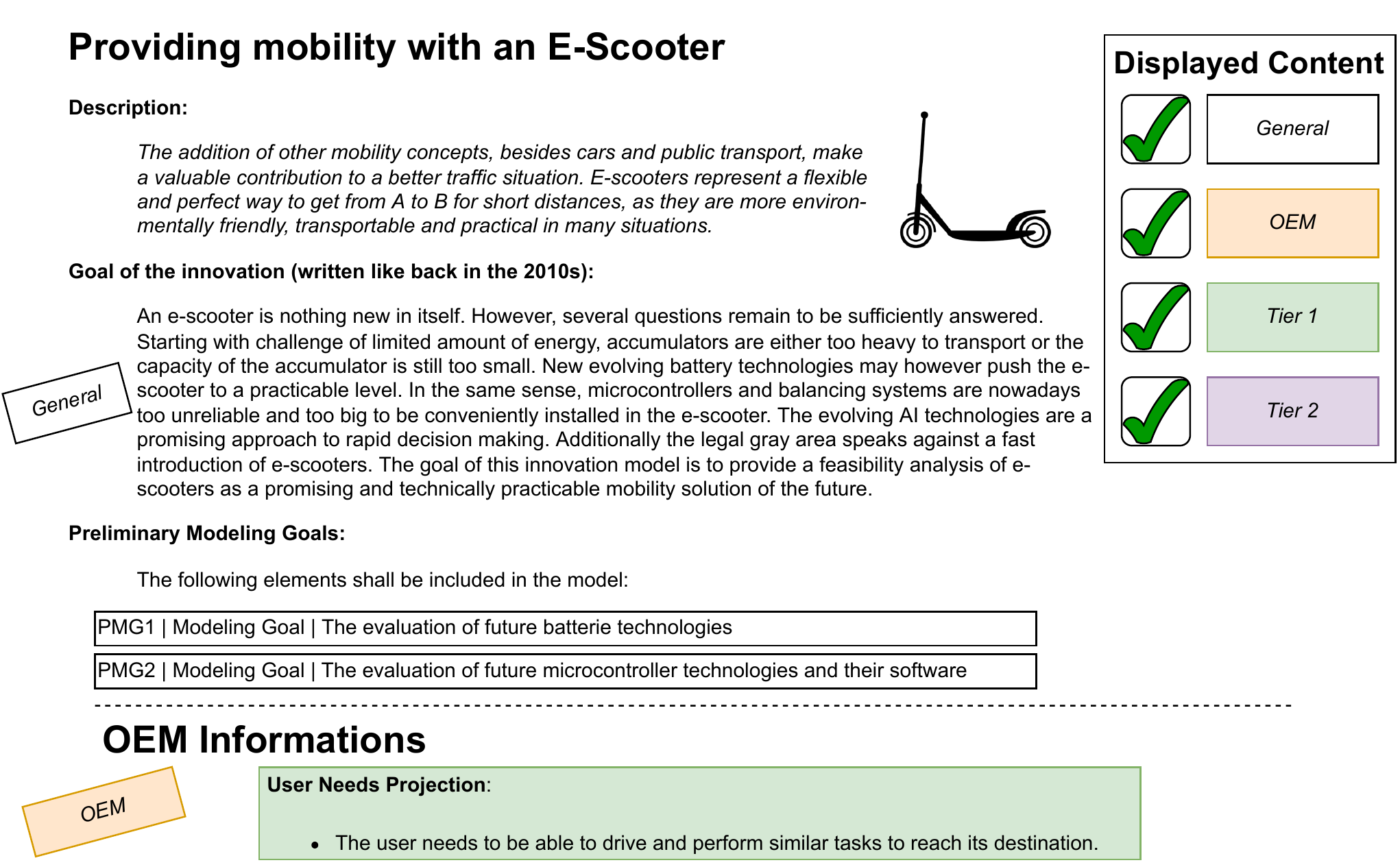}
	\caption{Strategy Perspective: part of the innovation description of the e-scooter.}
	\label{fig:imog:e-scooter:strategy}
\end{figure}

In the following, we illustrate the process steps with the innovation \enquote{Providing mobility with an e-scooter} (see Figure \ref{fig:imog:e-scooter:strategy}).
The Strategy Perspective of the e-scooter innovation includes a description with a vision and what the innovation is about, the goals written as text as well as goals listed as elements for cross referencing, information from the car manufacturer (OEM) regarding their estimated customer needs, their concern and possibly some additional bubble diagram for a better explanation of their interest and information from the other suppliers (Tier 1 and Tier 2) including their interest, diagrams, etc.

\subsubsection{Functional Perspective}
\label{sec:imog:methodology:fp}
The Functional Perspective focuses on the identification of the required features (end-user visible characteristics) and functions (traceable tasks or actions that a system shall perform) to design the innovation.
The features and functions of the Functional Perspective represent a derivative of the well-known feature models from \cite{kang1990feature}.
Optionally, User Stories or Use Cases can be created if the committee determines the need for more information on each feature and function.

\begin{figure}
	\includegraphics[width=\linewidth]{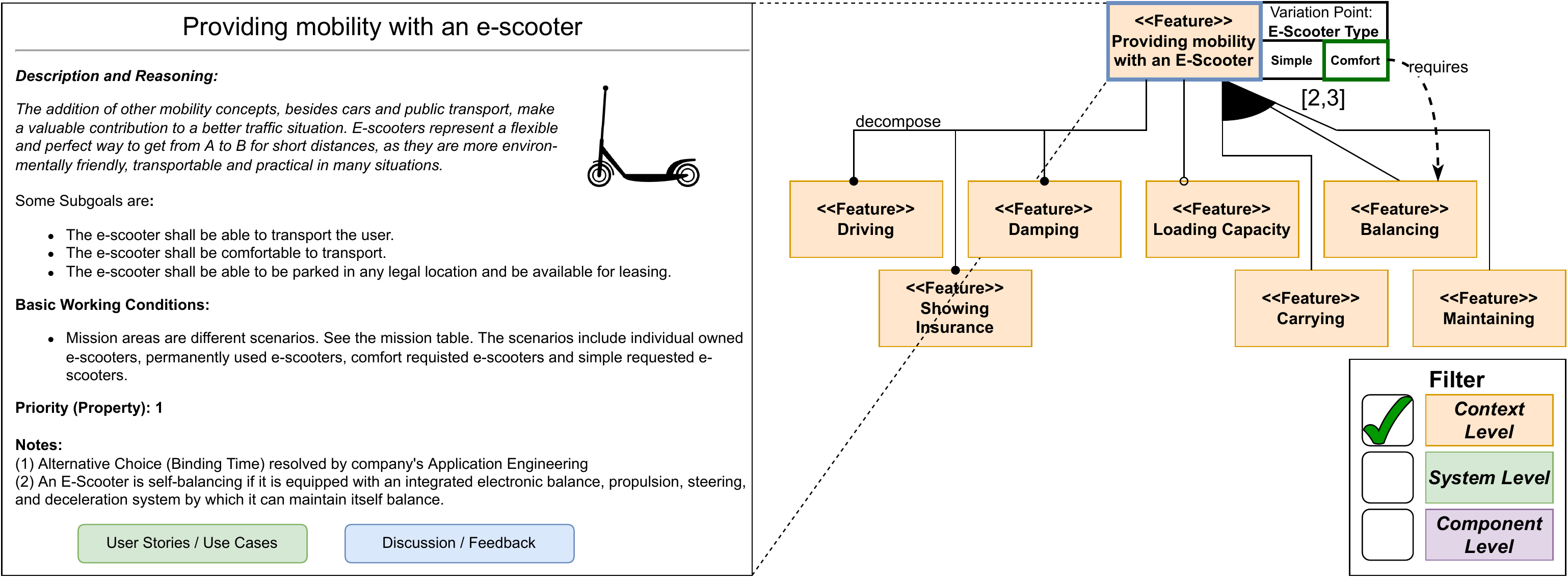}
	\caption{Functional Perspective: a part of the feature model of the e-scooter.}
	\label{fig:imog:e-scooter:fp}
\end{figure}

Considering the e-scooter example, the Functional Perspective model is partially depicted in Figure \ref{fig:imog:e-scooter:fp}.
It starts with \enquote{Providing mobility with an e-scooter} as its root feature, which is decomposed into several other features.
The mandatory relations are depicted by an arrow with a black circle as its head, and optional relations are depicted by an arrow with a white circle as its head.
Finally, an or-relation with cardinality is depicted as a black arc with several lines and a constraint relation (\enquote{requires}) as a line with a stereotype.
For more details about these described relations, see \cite{kang1990feature,czarnecki2000generative}. 
The Variation Point representation represents a labeled alternative relation, which has its own IMoG specific graphical depiction.
The e-scooter feature description can be seen on the left.
It includes a detailed textual description, aligned goals, basic working conditions and other properties, like notes, priorities or links to user stories and use cases.
The functions of the model are left out of the image.
One specialty of the Functional Perspective is the detailed description of each feature and function, which helps to understand what they actually represent.

\subsubsection{Quality Perspective}
\label{sec:imog:methodology:qp}
Based on the strategy description and the features and functions, the Quality Perspective captures quality requirements and constraints of each feature and function.
Requirement diagrams and requirement tables are adequate representations of the Quality Perspective.
These requirements and constraints, the features and functions as well as the strategy description encompass together the problem space.
The Quality Perspective does however contain another part:
the quality requirements and constraints of the solutions on the Structural Perspective.

\begin{figure}
	\includegraphics[width=\linewidth]{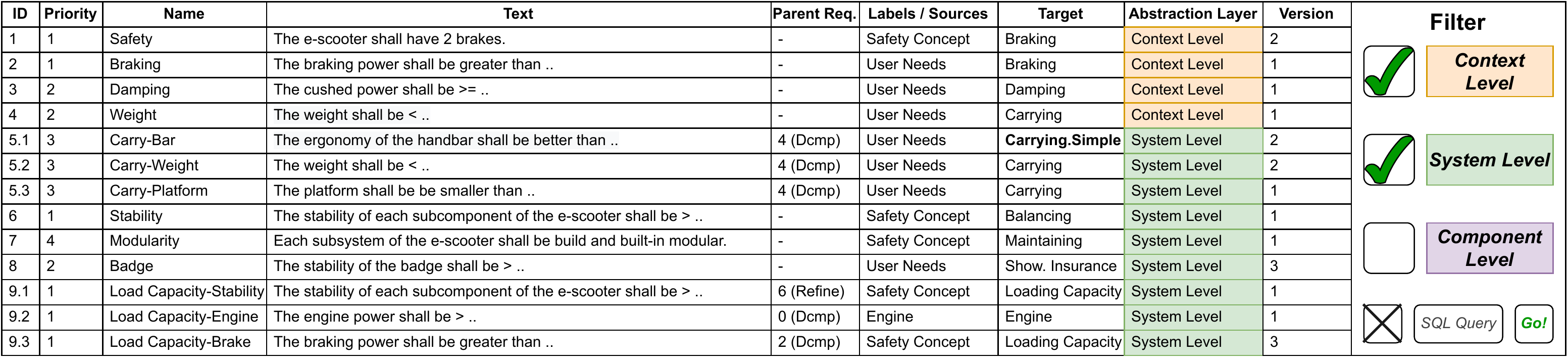}
	\caption{Quality Perspective: a table of requirements with many attributes, which reference features or functions of the Functional Perspective or solution blocks of the Structural Perspective.
	The details -- like the meaning of the attributes -- can be chosen individually depending on each innovation and are not further elaborated here.
	Filter functionality is of special importance for the Quality Perspective.}
	\label{fig:imog:e-scooter:qp}
\end{figure}

The Quality Perspective of the e-scooter innovation is depicted in Figure \ref{fig:imog:e-scooter:qp}.
It contains the quality requirements of the problem space and the solution space.

\subsubsection{Structural Perspective}
\label{sec:imog:methodology:structural}
\begin{figure}
	\centering
	\includegraphics[width=0.75\linewidth]{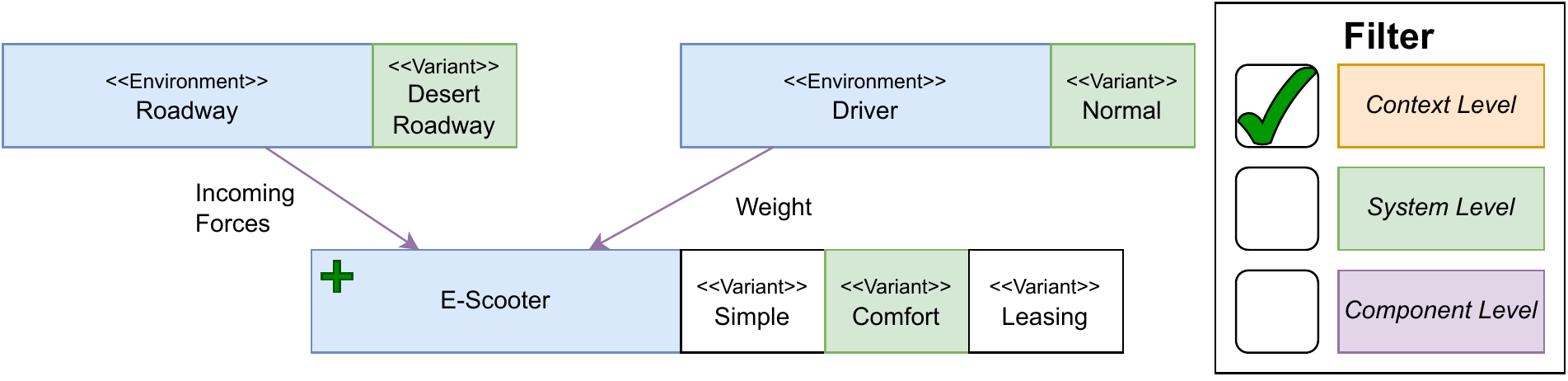}
	\caption{Structural Perspective - Context Level: A simple context model for the e-scooter.
	It contains the innovation (e-scooter) with the driver and roadway blocks (blue rectangles with a name and optionally a stereotype over the name).
	Each block has variants attached, that specify different forms of solutions.
	The variants are depicted as green and white boxes next to the solution blocks with the stereotype <<Variant>>.
	Furthermore, relations like \enquote{Incoming Forces} and \enquote{weight} are modeled as unidirectional purple arrows, where purple represents the color for relations stereotyped as <<effect>>.
	The solution blocks of the different abstraction levels are left out of the model.}
	\label{fig:imog:e-scooter:structural1}
\end{figure}

The Structural Perspective targets the modeling of the solution space.
It is worth mentioning, that the word \enquote{Structural} does not mean the relations of solution blocks to each other alone here, but it also includes properties and values of these solution blocks.
Starting with the environment on the context level, the interaction and effects between the environment and the innovation are modeled.
A simple environment description for the e-scooter may contain blocks for the street, the driver and the e-scooter (see Figure \ref{fig:imog:e-scooter:structural1}).
Each of them have variants attached, that specify different forms of solutions.
Each of these blocks own properties, which further refine them.
The added information is crucial for analyzing and evaluating the solution space.
Additionally, exemplary \enquote{Incoming Forces} and \enquote{Weight} effects are added.

\begin{figure}
	\includegraphics[width=\linewidth]{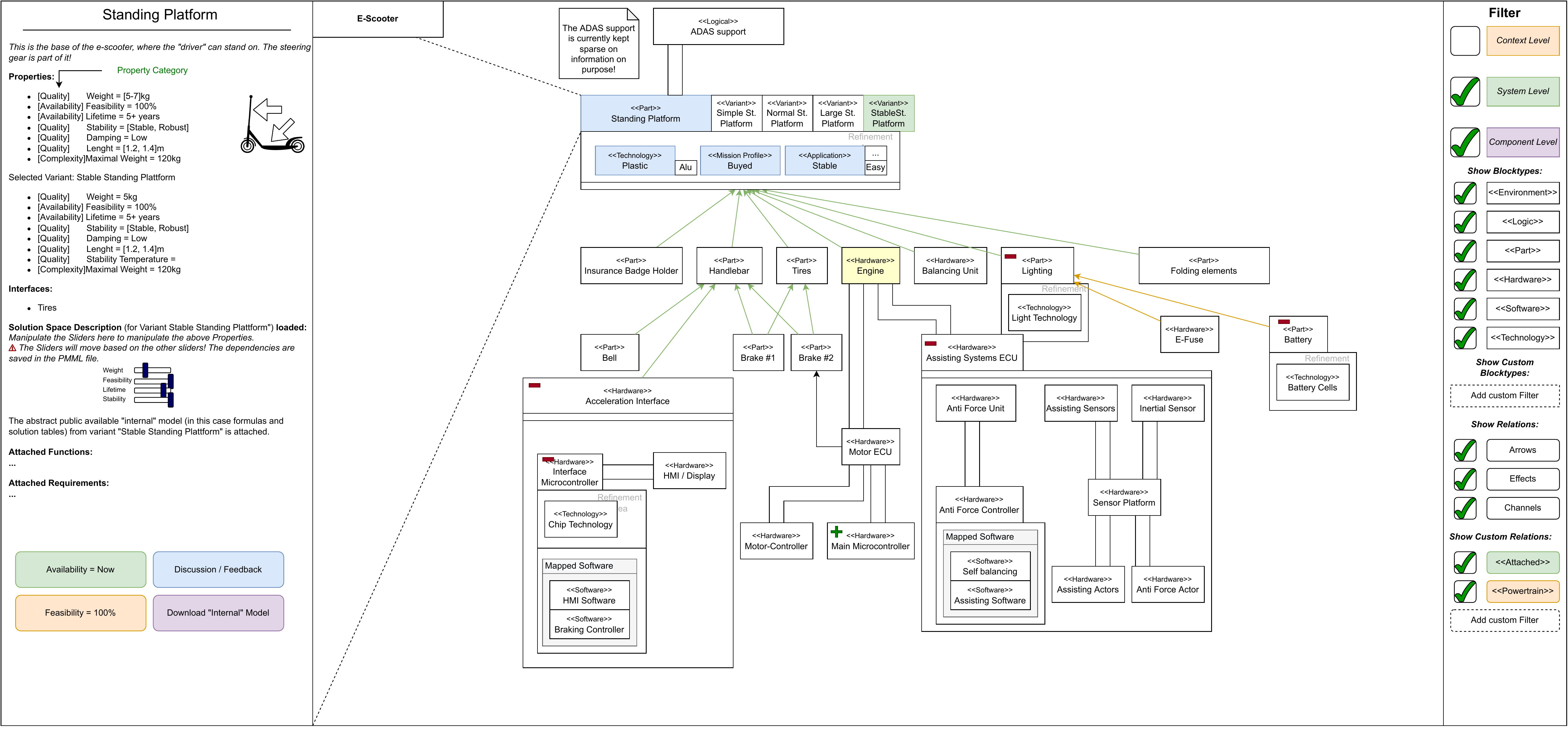}
	\caption{Structural Perspective - System Level:  The decomposition of the e-scooter into its system parts.
	It contains many blocks, variants, relations and channels (for modeling communication).
	This figure shall only give a glance at what may be included in the Structural Perspective.
	The exact details are out of scope of this paper.}
	\label{fig:imog:e-scooter:structural2}
\end{figure}

On the system level, the innovation is decomposed into components including software and hardware elements.
CPU, hardware architectures and mappings between them are part of the system level.
On the component level, these system blocks are decomposed into their atoms.
These may include sensor descriptions with parameters, functions, properties or abstract technologies.
The constraints and parameters of chosen technologies are particularly important on the component levels.
Additionally, requirements can be added to the solution blocks.
These requirements are added when creating the Structural Perspective, but are placed on the Quality Perspective and then referenced on the corresponding solution block on the Structural Perspective.
An example of a system model can be viewed in Figure \ref{fig:imog:e-scooter:structural2}:
it decomposes the e-scooter block known from Figure \ref{fig:imog:e-scooter:structural1} into several parts of the e-scooter.
The model elements are designed specifically for the microelectronic context.

\subsubsection{Knowledge Perspective}
\label{sec:imog:methodology:kp}
Based on the committee's insights from the problem and solution space, the reusable elements are stored in a database for later use in other innovation models.
This database is called Knowledge Perspective and builds the last perspective of IMoG.
The component database and knowledge base enable references made in future innovation models.
The database may, for example, contain sensor characteristics and constraints from road traffic regulations, with each element owning an id, a name, a type, an estimated year of availability and several properties depending on the context of innovation.

\subsubsection{Connecting Perspectives}
\label{sec:imog:methodology:connecting}
All perspectives were presented in detail.
However, their interconnection needs to be described.
These interconnections are already visible in Figure \ref{fig:imog1.4} and are described here shortly.
The elements of the Strategy Perspective can be referred by the features and functions, building the interconnection between the Strategy Perspective and Functional Perspective (represented by the <<references>> relations in Figure \ref{fig:imog1.4}).
The constraints are part of the Quality Perspective and own a target reference to the corresponding features and functions.
The same applies to the requirements, which are mapped on the Structural Perspective's solution blocks.
Thus the Quality Perspective has traces to both Functional Perspective and Structural Perspective (represented by the <<constrains>> relations in Figure \ref{fig:imog1.4}).
Each feature and function should be mapped on one or several solution blocks (represented by the <<allocate>> relations in Figure \ref{fig:imog1.4}).
This allocation is crucial, because it represents the interconnection of the problem space with the solution space.
Finally, there is the reference between the solution blocks of the Structural Perspective and the Knowledge Perspective  (represented by the <<references>> relations in Figure \ref{fig:imog1.4}).
Thus all perspectives are interconnected to each other.
Worth to note is, that the IMoG modeler must ensure that no inconsistencies arise (e.g., a requirement that is mapped on a feature or function, which is then allocated on a solution block that owns a contradicting requirement).

\begin{figure}
	\includegraphics[width=\linewidth]{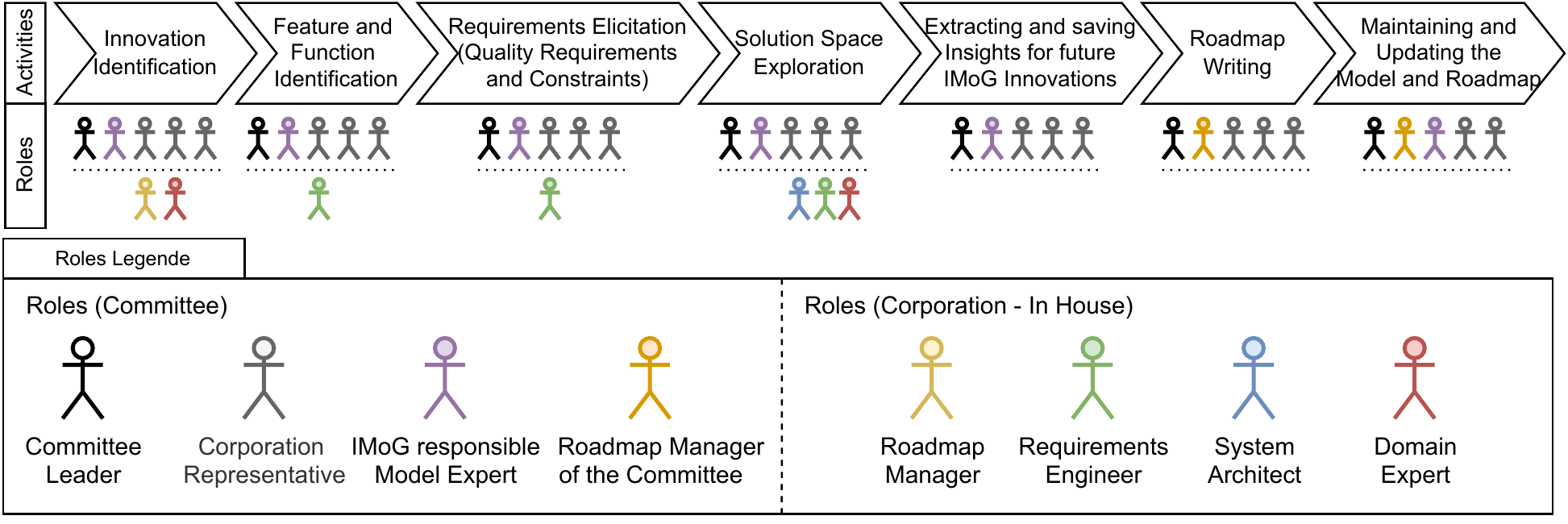}
	\caption{Activities (arrows) and roles of the working process. 
		The space between the activities represent nothing special and is for the sake of the graphical representation only.
		The roles of IMoG are divided into two groups: Committee roles and in-house employees of a corporation.}
	\label{fig:imog-process}
\end{figure}
\subsection{Process for IMoG}
\label{sec:imog:process}
Section \ref{sec:imog:methodology} described the IMoG methodology.
The questions how and when the methodology artifacts are used are not yet answered.
Therefore, this section describes IMoG's process to answer these questions. 
IMoG's process description contains the involved roles, their activities and the produced artifacts.
The roles of IMoG are divided into two groups (see Figure \ref{fig:imog-process}):
\begin{enumerate}
	\item The roles of the members of the automotive value chain committee. The roles are described in Table \ref{tab:roles:committee}.
	\item And the roles corporation employees, which specialize the role of the \enquote{Corporation Representative} to execute the specific activities of IMoG.
	These (in-house) employees help the committee by providing and compiling information.
	These roles are described in Table \ref{tab:roles:interns}.
\end{enumerate}

\begin{table}
	\caption{The involved roles in the automotive value chain committee}
	\begin{tabular}{p{0.22\textwidth}|p{0.75\textwidth}}
		\textbf{Roles} & \textbf{Description}\\\hline
		Committee Leader & The responsible person leading the roadmap committee.\\
		Corporation Representative & The responsible person of a corporation to coordinate the corporation internal tasks to produce the needed inputs for the roadmap. For the automotive value chain, this representative may be classified by OEM, Tier1 or Tier2.\\
		IMoG responsible Model Expert & The responsible person of creating and maintaining the IMoG model on the command of the committee members. The IMoG responsible Model Expert is not responsible for the content of the model (e.g. Shall A be decomposed to B and C? etc.) and only for the correct use of semantics and representation.\\
		Roadmap Manager of the committee & The roadmap manager of the committee is responsible for the creation and maintenance of the roadmap.
	\end{tabular}
	\label{tab:roles:committee}
\end{table}

\begin{table}
	\caption{The involved roles executing the required activities of IMoG}
	\begin{tabular}{p{0.22\textwidth}|p{0.75\textwidth}}
		\textbf{Roles} & \textbf{Description}\\\hline
		Roadmap Manager & The roadmap manager monitors the innovation status, reports to top management on the feasibility of the innovation, surveys new technologies from other partners, and updates the roadmap. The roadmap manager investigates trends and innovations. During innovation modeling, the roadmap manager performs the initial tasks and writes the roadmap after consulting with the other domain experts, requirements engineers, and system architects.\\
		Requirements Engineer & The requirements engineer creates initial top-level requirements for the innovation and captures them uniformly (formally or in natural language). The requirements engineer leverages the expertise of the domain experts and system architects to uniformly refine the requirements in the system models.\\
		System Architect & The system architect has the role of an interdisciplinary expert who designs systems by using modeling techniques. The system architect has know-how in the area of software - hardware design. In innovation modeling, the system architect takes on the role of the innovation modeler and its decomposition into subsystems.\\
		Domain Expert & The domain expert represents a specialist of a particular discipline covering subdomains of development. The domain expert supports the innovation modeling and evaluates its influences and dependencies of certain domain elements on other domain elements.
	\end{tabular}
	\label{tab:roles:interns}
\end{table}

We do not define a specific way through the IMoG matrix.
Instead, the members of the automotive value chain committee (and thus the two mentioned groups of roles) execute the following activities (see Figure \ref{fig:imog-process}):

The first step is called \textbf{Innovation Identification}.
This activity uses creative methods as well as market segment analysis to develop a new innovation idea and create an initial description.
The involved roles include the Committee Leader to set up and manage the meetings, the IMoG modeler responsible for creating the models and the corporation representatives in the roles of the Roadmap Manager for proposing their interests in the innovations as well as some Domain Experts for supporting the Roadmap Managers.
The filled Strategy Perspective constitutes the artifact of the \enquote{Innovation Identification} activity.

The second step is the \textbf{Feature and Function Identification}.
Its goal is to refine the problem understanding and create a feature hierarchy including optional User Stories and Use Cases based on the description of the innovation.
The involved roles include again the committee members of step 1 (Committee Leader, IMoG modeler and the corporation representatives).
In-house requirements engineers are involved in this step.
The filled Functional Perspective constitutes the artifact of the \enquote{Feature and Function Identification} activity.

The third step is the \textbf{Requirements Elicitation}, which adds quality requirements and constraints to the feature hierarchy and refines in this way the problem space further.
It is the last step focusing on the problem space.
The roles that are involved in this step are the same as in step 2.
The filled Quality Perspective constitutes the artifact of the \enquote{Requirements Elicitation (Quality Requirements and Constraints)} activity.

The solution space of the innovation is examined after the problem is sufficiently understood.
The corresponding step is called \textbf{Solution Space Exploration}.
It consists of modeling the possible solutions of the innovation with (sufficient) technical details.
The leader of this task is the system architect to examine and analyze the possible solutions.
The system architect gets support from the requirements engineer and the domain expert, however, their help is of supportive nature.
The filled Structural Perspective constitutes the artifact of the \enquote{Solution Space Exploration} activity.

After the solutions are examined, the committee extracts the insights gained by the created model and saves them in their database for further innovations.
This step is called \textbf{Extraction and Saving of the Insights}.
No in-house corporation role is needed.
The Knowledge Perspective constitutes the artifact of the \enquote{Extracting and saving Insights for future IMoG Innovations} activity.

The \textbf{roadmap writing} is the next step building upon the insights from the previous step.
The committee members meet up again and discuss the roadmap together.
The modeling activities are finished and the IMoG modeler is not needed for this task.
The roadmap manager takes responsibility for the roadmap writing, structures the document, and assigns tasks.
After this step, the main roadmapping activies are done.

Based on this roadmap, reoccurring meetings are established to \textbf{maintain and update} the roadmap.
The same roles are involved as in the writing of the roadmap.

It is not required to complete each of the seven steps before the next one is started (as usual in process models).
Instead, it is sufficient to draft each model of each step and refine them when necessary, similarly to what was proposed within the twin peaks model \cite{nuseibeh2001weaving}.

\section{Discussion and Lessons Learned}
\label{sec:evaluation}
We have consolidated IMoG by refining IMoG's perspectives and their corresponding model elements and by defining a suitable process for IMoG.
Furthermore, we applied IMoG on a few academic examples including the e-scooter example presented in this article.

In this section, we review the discussion of the evaluation of the journal article \cite{Fakih2021} and describe what we have learned from the examples.

The initial evaluation in the original proposal of IMoG~\cite{Fakih2021} stated two strengths: the appropriate level of abstraction for modeling innovations and the examined ways through the matrix.
The examined ways include, for example, a top-down diagonal approach from the Context Level of the Strategy Perspective down to the Component Level of the Structural Perspective or a bottom-up approach from the ideas of the semiconductor suppliers back to the context of the car manufacturers.

The appropriate level of abstraction was confirmed and further underlined by the examples: IMoG helped us to adequately tackle the innovations.
We reconsidered our opinion regarding the mentioned ways through the matrix.
Instead of specifying several possible ways through IMoG, we think it is rather appropriate to follow the mentioned process for IMoG presented in section \ref{sec:imog:process}.
Furthermore, iterating between the problem space and solution space perspectives similar to the process defined in the twin peaks model \cite{nuseibeh2001weaving} is in our opinion the most appropriate approach.

The initial evaluation in the original proposal of IMoG~\cite{Fakih2021} identified three potential limitations based on an academic example of wireless charging: scalability, detailed behavioral models, and bridging to product level models.
The application of IMoG to the larger example of the e-scooter sheds more light onto these topics. 
First, we did not encounter any issues regarding scalability in the example modeled here, which indicates that IMoG as such does not introduce unnecessary and unmanageable complexities.
Second, our example here confirms the view that the absence of detailed behavioral models is actually a strength: Details are not required and should be left out in abstract innovation modeling.
Nonetheless, such detailed models should be possible to be attached to solution blocks whenever needed.
Finally, the bridge between an IMoG model to a product level model remains properly solvable:
Bridging the gap by referring IMoG's elements, using transformations of IMoG models to established system level development languages or by translating the IMoG model into a development focused framework (see Broy et al. in \cite{broy2009toward}) with adding the behavioral aspects to the designed framework are the recommended choices.

While applying IMoG to the examples we learned two more lessons:
Reordering the perspectives into the problem space and solution space made it easier to apply IMoG.
This distinction got added to the design principles of IMoG (see section \ref{sec:imog:overview}).
Another lesson was, that interpreting abstraction levels as filter functionality is better suited for the modeler than interpreting abstraction levels as a division into diagrams.
We examined that the division of an innovation model into several pieces would do more harm regarding its user experience and usefulness than it would help.
\section{Conclusion}
\label{sec:conclusion}
This article presents the modeling methodology \enquote{Innovation Modeling Grid} (IMoG).
IMoG targets the creation and maintenance of a public roadmap by an automotive value chain committee.
A particular focus of IMoG is the tailoring to an adequate level of abstraction for modeling innovations.
The methodology IMoG focuses on splitting its models into the well known problem space and the solution space.
Half of its five perspectives cover the problem space and the other half focus on the solution space.
Each perspective covers a different aspect of an innovation.
Additionally, IMoG contains abstraction levels and builds -- together with the perspectives -- a matrix.

We presented the design principles of IMoG, evaluated the methodology with the larger innovation example of an e-scooter and introduced a process for IMoG.
We revisited the strengths and the limitations of IMoG from our preliminary evaluation and described our lessons learned.
Overall, we see a high potential in the IMoG methodology.
Our next steps include an evaluation of IMoG in an industrial context.
To this end, we are currently implementing a dedicated tooling for IMoG.

\subsubsection{Acknowledgments}
This work has been supported by the GENIAL! project as funded by the German Federal Ministry of Education and Research (BMBF) under the funding code 16ES0865-16ES0876 in the ICT 2020 funding programme.

\bibliographystyle{splncs04}
\bibliography{bibliography}
\end{document}